\documentclass[aps,prb,superscriptaddress,sd,reprint]{revtex4-1}
\usepackage{graphicx}
\usepackage{dcolumn}
\usepackage{bm}
\usepackage[colorlinks,linkcolor=blue,citecolor=blue,urlcolor=blue]{hyperref}
\usepackage{amsmath,amsthm,amssymb}
\usepackage{amsfonts}
\usepackage{icomma}
\usepackage{placeins}
\usepackage{color}
\usepackage[separate-uncertainty = true, exponent-product = \times]{siunitx}
\DeclareSIUnit\torr{Torr}

\makeatletter
\let\cat@comma@active\@empty
\makeatother

\begin{document}
 
\title{Spin-Orbit Torques in ferrimagnetic GdFeCo Alloys}

\author{Niklas Roschewsky}
\affiliation{Department of Physics, University of California, Berkeley, California 94720, USA}

\author{Tomoya Matsumura}       
\affiliation{Department of Electrical Engineering and Computer Science, Nagoya University, Nagoya 464-8603, Japan}

\author{Suraj Cheema}
\affiliation{Department of Materials Science and Engineering, University of California, Berkeley, California 94720, USA}

\author{Frances Hellman}
\affiliation{Department of Physics, University of California, Berkeley, California 94720, USA}
\affiliation{Materials Science Division, Lawrence Berkeley National Laboratory, Berkeley, California 94720, USA}

 \author{Takeshi Kato}
\affiliation{Department of Electrical Engineering and Computer Science, Nagoya University, Nagoya 464-8603, Japan}

\author{Satoshi Iwata}
\affiliation{Institute of Materials and Systems for Sustainability, Nagoya University, Nagoya 464-8603, Japan}

\author{Sayeef Salahuddin}
\email{sayeef@berkeley.edu}
\affiliation{Materials Science Division, Lawrence Berkeley National Laboratory, Berkeley, California 94720, USA}
\affiliation{Department of Electrical Engineering and Computer Sciences, University of California, Berkeley, California 94720, USA}

\date{\today}

\begin{abstract}
The spin-orbit torque switching of ferrimagnetic Gd$_x$(Fe$_{90}$Co$_{10}$)$_{100-x}$ films was studied for both transition metal (TM)-rich and rare earth (RE)-rich configurations.  The spin-orbit torque driven magnetization switching follows the same handedness in TM-rich and RE-rich samples with respect to the total magnetization, but the handedness of the switching is reversed with respect to the TM magnetization. This indicates that the sign of the spin-orbit-torque-driven magnetic switching follows the total magnetization, although transport based techniques such as anomalous Hall effect are only sensitive to the transition metal magnetization. These results provide important insight into the physics of spin angular momentum transfer in materials with antiferromagnetically coupled sublattices.
\end{abstract}

\maketitle

Magnetization dynamics at interfaces has been investigated extensively over the last three decades \cite{PhysRevLett.55.1790,Slonczewski1996L1,PhysRevB.54.9353}. In that context `spin-orbit torque' (SOT) has  received a lot of interest recently. Here, a charge current in a heavy metal is converted into a spin current via spin-orbit coupling and injected into an adjacent ferromagnet~\cite{PhysRevLett.109.096602,Liu04052012,Miron2011,MihaiMiron2010}. The transfer of angular momentum from the spins to the ferromagnet causes a torque on the magnetization which can switch the magnet~\cite{Miron2011,MihaiMiron2010}. To date, most studies of SOT have concentrated on $3d$ ferromagnets such as Co~\cite{PhysRevLett.109.096602,Miron2011,:/content/aip/journal/apl/107/23/10.1063/1.4937443,:/content/aip/journal/apl/105/21/10.1063/1.4902529} FeCo~\cite{Emori2013a}, FeCoB~\cite{PhysRevB.89.144425,Qiu2015,Fan2014}, FePd~\cite{Lee2014} or transition metal rich ferrimagnetic alloys such as TbFeCo \cite{:/content/aip/journal/apl/106/13/10.1063/1.4916665}. Here we report SOT switching of ferrimagnetic GdFeCo alloys with both rare earth (RE) rich or transition metal (TM) rich configurations with bulk perpendicular magnetic anisotropy (PMA) at room temperature. 

Our study is based on Gd$_x$(Fe$_{90}$Co$_{10}$)$_{100-x}$ thin films, where the antiferromagnetic ordering between Gd $4f$ and FeCo $3d$ magnetic moments is mediated by Gd 5d electrons via $4f$-$5d$ exchange interaction and $3d$-$5d$ hybridization~\cite{Brooks1991,Pogorily2005} and indirect RKKY conduction band exchange. The magnetic properties of Gd$_x$(Fe$_{90}$Co$_{10}$)$_{100-x}$ can be varied by changing the composition $x$. A comparison between TM-rich and RE-rich samples allows us to conclude that the anomalous Hall effect (AHE) changes sign for these two different samples, in accordance with literature~\cite{Asomoza1977,Malmhall1983,Mimura1976,Shirakawa1976}, while the SOT-driven magnetic switching has the same sign in both samples. Further, the effective magnetic fields, induced by damping-like and field-like SOT, do not show any change in the sign.
\begin{figure}
\includegraphics[bb = 0 0 241 170]{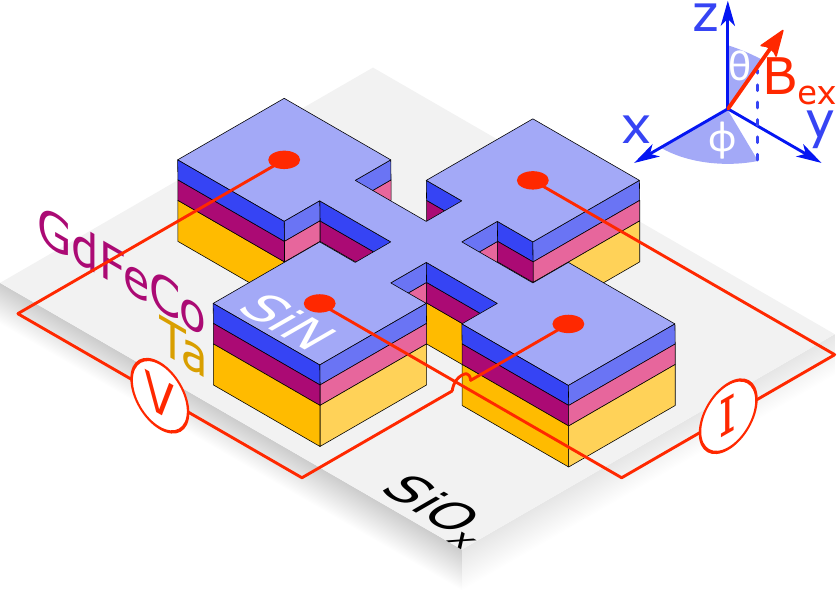}
\caption{\label{fig:Setup} Schematic of the measurement geometry used for SOT measurements. The samples are Ta/$a$-GdFeCo/$a$-SiN films, where the $a$-GdFeCo layer exhibits perpendicular magnetic anisotropy. Films are patterned into Hall bar structures for transport measurements. A current is applied along the $x$-direction while the anomalous Hall effect voltage is measured along the $y$-direction. An external magnetic field can be applied to the sample.}
\end{figure}

Ta(10)/Gd$_x$(Fe$_{90}$Co$_{10}$)$_{100-x}$(5)/SiN(5) films (thickness in nm) were grown by RF magnetron sputtering on thermally oxidized silicon substrates with compositions $x=21$ (TM-rich) and $x=28$ (RE-rich). The base pressure during deposition was lower than \SI{1e-8}{\torr}. The SiN overlayer is used to prevent oxidation. After growth, PMA was confirmed with magnetometry. Hall bar mesa structures with a width of \SI{20}{\micro\metre} were patterned using optical lithography and ion milling. The layout of the sample and the measurement setup are shown in Fig.~\ref{fig:Setup}. DC or AC currents are applied along the $x$-direction while the transverse voltage is measured. The orientation of the external magnetic field $B_\text{ex}$ is defined by spherical coordinates $\phi$ and $\theta$. This device structure is used for AHE and planar Hall effect (PHE) measurements. We find that the resistivity of our samples is $\rho_\text{TM}=\SI{320}{\micro\ohm\centi\meter}$ and $\rho_\text{RE}=\SI{342}{\micro\ohm\centi\meter}$ for the TM-rich and the RE-rich sample respectively.
\begin{figure}
\includegraphics[bb = 0 0 241 172]{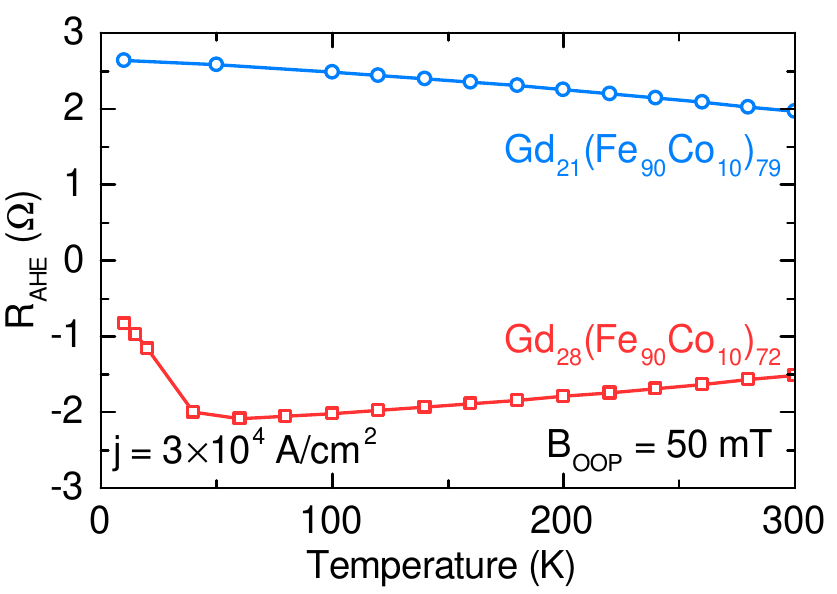}
\caption{\label{fig:figure2} The anomalous Hall resistance as a function of temperature is plotted for the TM-rich sample (blue circles) and the RE-rich sample (red squares). The measurements are taken with an applied out-of-plane field $B_\text{OOP}$ of \SI{50}{\milli\tesla}. The investigated samples show no compensation point below room temperature.}
\end{figure}

The AHE of GdFeCo as a function of temperature is shown in Fig.~\ref{fig:figure2}. An out-of-plane magnetic field of \SI{50}{\milli\tesla} is applied to fix the magnetization and prevent domain nucleation. The AHE resistance is proportional to the out-of-plane component of the TM magnetization~\cite{Asomoza1977,Malmhall1983,Mimura1976,Shirakawa1976}. In the RE-rich sample, the transition metal moment is aligned antiparallel to the external magnetic field and thus a negative AHE resistance is measured. Since the AHE resistance does not change sign over the whole measurement range, there is no compensation point for the magnetization in either our samples. The small decrease in AHE resistance ($R_\text{AHE}$) as a function of temperature, seen in both samples, is due to a decrease of the saturation magnetization in the TM. The decrease of $R_\text{AHE}$ at temperature below \SI{50}{\kelvin} in the RE-rich sample is due to  a change of the magnetic anisotropy where the easy axis tilts from out-of-plane to in-plane. This is because the net magnetization of the sample has grown large (due to Gd moment increasing more rapidly than TM moment with decreasing temperature) causing dipolar coupling energy to become larger than perpendicular anisotropy energy.

After confirming that the magnetization of Gd$_x$(Fe$_{90}$Co$_{10}$)$_{100-x}$ is dominated by the TM magnetic moment for $x=21$ and by the RE magnetic moment for $x=28$, we concentrate on room temperature measurements. 
\begin{figure}
\includegraphics[bb = 0 0 241 246]{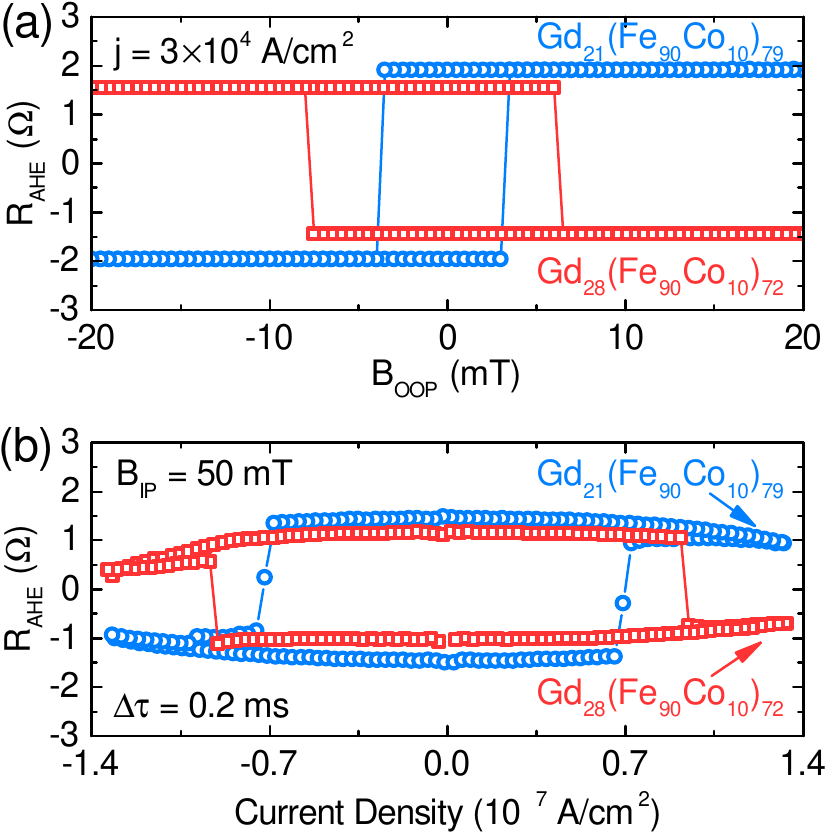}
\caption{\label{fig:figure3} \textbf{(a)} Anomalous Hall resistance at room temperature as a function of out-of-plane magnetic field $B_\text{OOP}$. The sign of the anomalous Hall effect is different in the TM-rich and RE-rich samples. \textbf{(b)} Spin-orbit-torque-driven switching of the magnetization. A magnetic field of \SI{50}{\milli\tesla} is applied parallel to the current direction to break the symmetry. The current pulse width is $\Delta\tau=\SI{200}{\micro\second}$. The sign of the SOT-driven switching is the same in both samples.}
\end{figure}
Figure~\ref{fig:figure3}(a) shows the anomalous Hall voltage as a function of the external magnetic field applied along the easy axis of the magnet ($\theta=0$). The current density is small to avoid heating- and SOT-induced effects. Sharp switching in this measurement indicates good PMA. The RE-rich sample shows a higher coercive field because it is closer to the compensation temperature. As mentioned earlier, the AHE has opposite sign for the TM-rich sample and RE-rich sample because it is proportional to the out-of-plane component of the TM. This leads to the different handedness of the hysteresis loops. 

The SOT-induced switching of the magnetization is shown in Fig.~\ref{fig:figure3}(b). During the measurement a \SI{50}{\milli\tesla} in-plane magnetic field with $\phi=0$ and $\theta=\pi/2$ is applied to break the symmetry. \SI{200}{\micro\second} long current pulses are used to switch the magnetization. Consider the TM-rich sample first: If we call the negative AHE resistance state down (because that is the equilibrium state if a large negative magnetic field is applied along the easy axis), then a positive current pulse will switch the total magnetization from down to up. In the RE-rich sample we call the positive AHE resistance state down (again, this is the equilibrium state if a negative magnetic field is applied along the easy axis). A positive current pulse will then switch the total magnetization of the RE-rich sample from down to up, \textit{just as it did for TM-rich sample}. Thus, SOT-driven magnetic switching follows the same handedness in both samples with respect to the total magnetization (down $\rightarrow$ up for positive current pulses). However, SOT-driven switching has a different sign in TM- and RE-rich samples with respect to the FeCo magnetization as measured by the AHE and thus the handedness of the hysteresis is different in both samples as shown in Fig.~\ref{fig:figure3}(b).

Next we performed harmonic Hall voltage measurements following Hayashi \textit{et al.}~\cite{PhysRevB.89.144425} to characterize the effective magnetic fields induced by SOT. To this end, an AC current with $\omega=\SI{97}{\hertz}$ is applied to the sample and the first- and second harmonic voltage responses ($V_\omega$ and $V_{2\omega}$) are measured. To measure the Slonczewski-like SOT $H_\text{SL}$, an external magnetic field is applied parallel to the current direction, while the magnetic field is applied perpendicular to the current to measure the field-like SOT $H_\text{FL}$. The field is swept quasistatically between $\pm\SI{60}{\milli\tesla}$. During the measurement, a small out-of-plane field ($\approx\SI{5}{\milli\tesla}$) is applied to prevent the magnet from breaking into domains. The effective fields can be calculated using the following procedure:
\begin{align}
H'_\text{SL,FL}&=\Bigg(\frac{\partial V_{2\omega}}{\partial H_\text{SL,FL}}\Bigg)\times\Bigg(\frac{\partial^2 V_{\omega}}{\partial H^2_\text{SL,FL}}\Bigg)^{-1}\, ,\\
H_\text{SL,FL}&=-2 \frac{H'_\text{SL,FL}\pm 2 \xi H'_\text{FL,SL}}{1-4\xi^2}\label{eq:EffectiveField}\, .
\end{align}
Here $V_\omega$ and $V_{2\omega}$ are the first and second harmonic Hall voltages, respectively. The plus (minus) sign in eqn.~\eqref{eq:EffectiveField} applies to $m$ pointing along the positive (negative) direction. $\xi$ is the ratio of planar Hall effect and anomalous Hall effect: $\xi=\Delta R_\text{PHE}/\Delta R_\text{AHE}$. In our samples we find $\xi_\text{TM}=0.072$ for the TM-rich sample and $\xi_\text{RE}=0.069$ for the RE-rich sample, measured via in-plane rotations of the magnetic field: $R(\phi)$ at $\theta=\pi/2$. 
\begin{figure}
\includegraphics[bb = 0 0 241 246]{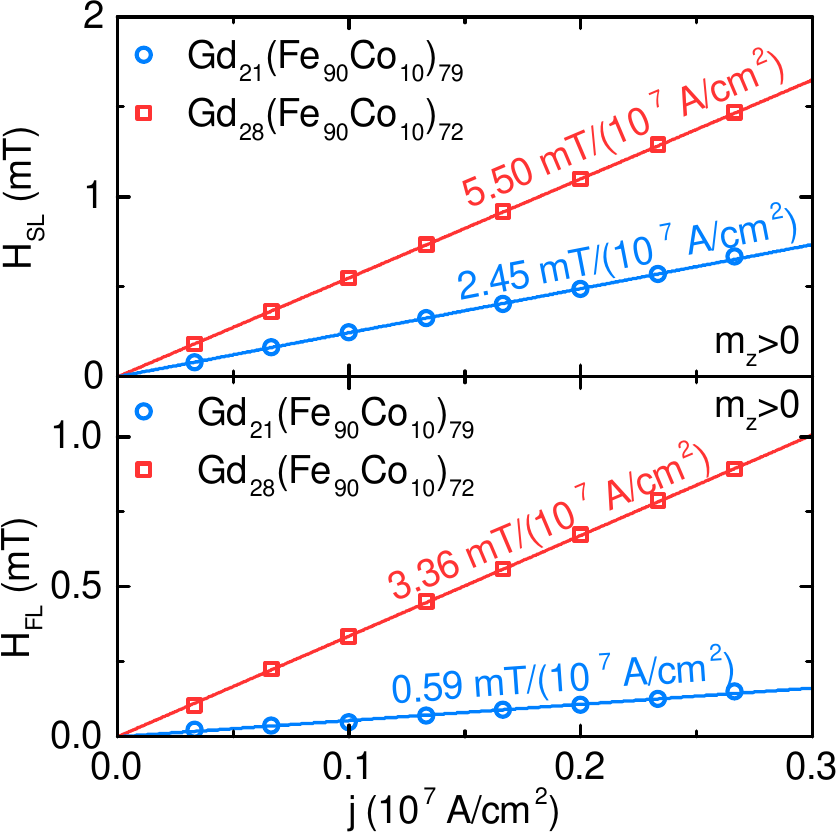}
\caption{\label{fig:figure4} The upper panel shows the longitudinal effective field as a function of the current density in the Ta layer for the TM-rich and the RE-rich sample while the lower panel shows the transverse effective field. The solid lines are fits to the experimental data. All measurements were taken with an out-of-plane field of $\approx\SI{5}{\milli\tesla}$ to ensure $m_z>0$.}
\end{figure}

The effective damping field is shown in the upper panel of Fig.~\ref{fig:figure4}. For both samples, the fields are a linear function of the current density. This linear relationship also shows that heating does not play a role in these measurements. The data was recorded for $m$ pointing along the positive $z$-direction. However it was confirmed that the Slonczewski field changes sign for $m$ pointing along the negative $z$-direction. The effective fields for the TM-rich and the RE-rich sample have the same sign, which is in agreement with the switching measurements reported earlier. It is important to notice that the effective damping field in the RE-rich sample is twice as large in magnitude as in the TM-rich sample at a given current density. The lower panel of Fig.~\ref{fig:figure4} shows the field-like (FL) torque. The field-like torque follows the same trend in that it is smaller for the TM-rich sample. It was confirmed that the FL-torque does not change sign for $m_z<0$ for both samples.

Mechanisms for the anomalous Hall effect discussed in literature include Berry phase~\cite{Onoda2002,Karplus1965,Baily2005}, side-jump scattering~\cite{Berger1964} and screw scattering~\cite{Smit1955,Nagaosa2010}. This indicates that the AHE is ultimately a conduction electron effect. According to Hund's rules Gd has a large magnetic moment due to a half filled $4f$-shell; the f-shell is localized and does not contribute to conduction (except by scattering of conduction electrons). In Fe, however, the d-band is spin-split and thus the conduction electrons are spin polarized. For that reason we expect the AHE to be dominated by the TM conduction electrons and consequently to follow the sign of $m_\text{TM}$ as seen in the experiments described above.

SOT, on the other hand, is an angular momentum transfer process. Due to a current in the Ta, electrons of one spin species will diffuse into the adjacent GdFeCo layer. This spins lie along the $y$-axis and are not collinear to the magnetization vector in the magnet and thus they are not eigenstates. Therefore they will begin to precess around the magnetization axis and average out within the spin coherence length~\cite{Kovalev2002}. If angular momentum is conserved in the system, there needs to be an angular momentum transfer process from the injected electrons to the magnetization of the ferromagnet. The change of the angular momentum (eg. the torque) is of the form  $\bm{\tau}=\tau_0\, \bm{m}\times(\bm{\sigma}\times \bm{m})$, where $\bm{\sigma}$ is the spin polarization~\cite{Brataas2012}.

Given that the antiferromagnetic coupling between the Gd $4f$ and the FeCo $3d$ magnetic moments is mediated via Gd $5d$ states, it would be reasonable to assume that the electrons diffusing into the GdFeCo due to the presence of the SHE also couple antiferromagnetically to the Gd moments. This would mean the SOT on Gd and FeCo has a different sign, in contradiction with our experiments. Therefore we conclude that the angular momentum transfer process to the Gd is ``ferromagnetic'', similar to the usual angular momentum transfer process in a $3d$ magnet. The result is consistent with spin torque measurements in GMR spin vales, where the sign of the GMR effect changes for TM- and RE-rich samples, but the sign for spin-torque switching remains unchanged~\cite{Jiang2006}.

It should be noted that in addition to a magnetization compensation point, an angular momentum compensation point has been observed in GdFeCo thin films due to the different gyromagnetic ratios of the Gd and FeCo sublattices~\cite{Stanciu2006}. However, angular momentum compensation does not play a role in our experiments with the RE-rich sample as the angular momentum compensation temperature is above the magnetization compensation temperature. The TM-rich sample  is far away from compensation.  

In conclusion we have studied SOT in Ta/GdFeCo/SiN structures with PMA. It was shown by temperature dependent measurements that the AHE is proportional to the out-of-plane magnetization of the transition metal. The spin-orbit-torque-driven magnetization switching follows the same handedness in TM-rich and RE-rich samples with respect to the total magnetization, but the handedness of the switching is reversed with respect the to TM magnetization. Therefore, we conclude that the angular momentum transfer process due to SHE in Ta has the same sign in both magnetic sub-systems, FeCo and Gd. In addition, $2\omega$ measurements confirm that the effective fields induced by spin-orbit torque have the same sign in both samples. However, the effective field induced in the RE-rich sample is twice as large as in the TM-rich sample, even bigger than effective fields observed in Ta/CoFeB samples. 

We thank Johannes Mendil, Pietro Gambardella and Dominic Labanowski for fruitful discussions. Research was supported by the Director, Office of Science, Office of Basic Energy Sciences, Materials Science and Engineering Division and the U.S. Department of Energy under Contract No. DE-AC02-05-CH11231 within the NEMM program (KC2204).  Device fabrication was supported by the STARNET/FAME Center.


\FloatBarrier

\bibliography{literature}

\end{document}